\documentclass[amsmath, amsfonts, superscriptaddress, showpacs, twocolumn, prb]{revtex4-1}
\usepackage{graphicx}
\usepackage{epsfig}
\usepackage{bm}
\usepackage{dcolumn}
\usepackage{amsmath}
\usepackage{amssymb}

\begin{document}

\title{Decay of the Kohn mode in hydrodynamic regime}

\author{A.~Iqbal}
\affiliation{Department of Physics and Astronomy, University of Iowa, Iowa City, Iowa 52242, USA}

\author{A.~Levchenko}
\affiliation{Department of Physics and Astronomy, Michigan State
University, East Lansing, Michigan 48824, USA}

\author{M.~Khodas}
\affiliation{Department of Physics and Astronomy, University of Iowa, Iowa City, Iowa 52242, USA}
\affiliation{Racah Institute of Physics, Hebrew University of Jerusalem, Jerusalem 91904, Israel}

\begin{abstract}
We develop a hydrodynamic description of the collective modes of interacting liquids in a quasi-one-dimensional confining potential. By solving Navier-Stokes equations we determine analytically excitation spectrum of sloshing oscillations. For parabolic confinement, the lowest frequency eigenmode is not renormalized by interactions and is protected from decay by the Kohn's theorem, which states that center of mass motion decouples from internal dynamics.  We find that the combined effect of potential anharmonicity and interactions results in the depolarization shift and final lifetime of the Kohn mode. All other excited modes of sloshing oscillations thermalize with the parametrically faster rate. Our results are significant for the interpretation of recent experiments with trapped Fermi gases observed weak violation of the Kohn theorem.     
\end{abstract}



\maketitle

\textit{Introduction}. Properties of quantum liquids in one-dimension (1D), as realized experimentally in nanoscale semiconducting wires, carbon nanotubes, laser traps of cold atoms, and edge channels of the quantum Hall effect, continue to attract tremendous attention in the physics research (see Refs.~\cite{Review-1,Review-2,Review-3} for recent reviews and references herein).  With increasing sophistication in high precision measurements and techniques these systems provide serious tests for the existing theoretical models, such as Luttinger liquid theory~\cite{Haldane,Stone,Gogolin,Giamarchi}, and ultimately challenge their completeness.  For example, the powerful approach of the Luttinger liquid formalism allows to account for the interaction effects nonperturbatively. However, this model does not adequately describe relaxation phenomena due to built-in approximation of the linearized quasiparticle dispersion, which by virtue of the kinematic constraints effectively closes the phase space available for inelastic scattering. In certain special cases, the lack of relaxation may be a generic property of the many-body system because of its complete integrability \cite{Mattis,Sutherland}. Alternatively, vanishing relaxation rates may happen because of the reasons prescribed by the Kohn theorem~\cite{Kohn,Dobson}.     

Motivated by recent experiments \cite{Kinast,Bartenstein,Altmeyer,Wright} we study relaxation of collective excitations in interacting two-dimensional (2D) systems confined along one of the two spacial dimensions. These systems interpolate between strictly 1D limit of Luttinger liquids and the two-dimensional Fermi liquids. The geometrical confinement in such systems splits the single band spectrum into multiple 1D subbands. The convenient and practically justified model idealization is the interacting particles confined by a harmonic potential. In this case the 1D subbands are equidistantly separated by a frequency of oscillations $\omega_\perp$ across the channel. Similar to the spectrum linearization in the strictly 1D liquids, harmonic approximation in the quasi-1D case on one hand simplifies the dynamics, and at the same time does not allow for proper description of thermalization processes. One necessarily has to account for the confinement anharmonicity, which thermalizes the motion across the channel in much the same way as the spectrum nonlinearity causes the relaxation of charge and spin excitations in 1D quantum wires \cite{Khodas,Barak,Karzig,Micklitz,Levchenko}. 

Despite the similarity with 1D, the relaxation of transversal excitations has a few distinct features setting these two problems apart. The kinematical constrains of momentum and energy conservations operational in 1D are less restrictive in quasi-1D. In contrast to the 1D case, which require three-particle scattering processes, the two-body collisions do cause the relaxation via the inter-subband transitions. Thermalization in quasi-1D may nevertheless be prohibited due to the Kohn theorem rather than kinematical restrictions. 

This theorem states that the motion of the system as a whole is unaffected by interactions.
Classically, it follows as the translationally invariant interaction energy is insensitive to the system displacement as a whole. For the same reason, quantum mechanically, the interaction drops out of the center of mass Heisenberg equation of motion \cite{Dobson,Brey}. In a quantum Fermi liquid the Kohn theorem follows from the Landau Galilean invariance relation between the quasiparticle effective mass and the first angular harmonic of the interaction amplitude \cite{Iqbal}.

In all of the above cases the Kohn theorem states that if the confining potential is harmonic the collective sloshing oscillations proceed without decay. The frequency of the Kohn, or so called sloshing mode, $\omega_\perp$, is insensitive to interaction, temperature and particle statistics \cite{Dobson,Brey,Iqbal}. 

This fact makes the observation of the Kohn mode possible in a wide variety of systems. In semiconductor quantum wires the Kohn mode is observed in optical transmission at far infrared \cite{Drexler,Wendler}. In trapped ultracold Fermi gas of $^6$Li the Kohn mode of a half KHz frequency was excited by sudden displacement of the trap and detected by absorption imaging of a released cloud \cite{Altmeyer,Pantel}. 

The weak violation of the Kohn theorem due to anharmonicity observed in the above three classes of systems plays a key role in the data interpretation.  In the case of the semiconductor quantum wires it controls the line broadening and the higher harmonics of light transmission \cite{Drexler}. 
The observed sloshing frequency of an atomic cloud in the optical trap shows systematic deviations from the Kohn theorem prediction  \cite{Riedl}. Such deviations grow with heating as the expanding atomic cloud senses a progressively less parabolic confining potential. Here we concentrate on two fundamental aspects of Kohn theorem violation: (i) depolarization shift of the sloshing frequency, and (ii) final lifetime of sloshing oscillations. We approach this problem based on a very general grounds of hydrodynamic theory, which accurately describes most liquids at length scales long compared to the particle-particle mean-free path.

\begin{figure}
\includegraphics[width=0.9\columnwidth]{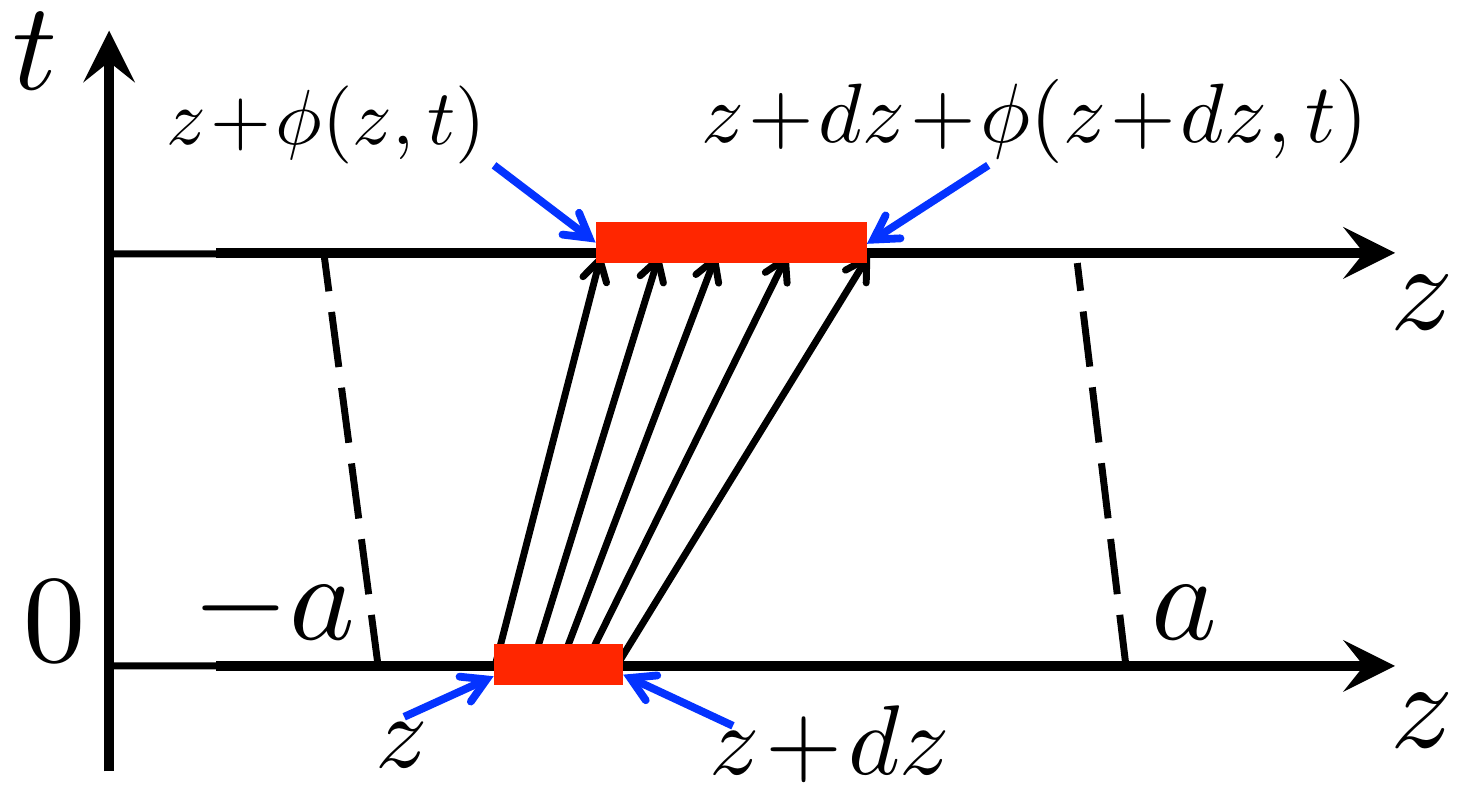}\\
\caption{(color online)
The definition of the displacement field $\phi(z,t)$ in Lagrangian formulation. 
At the time, shown as vertical axis, $t=0$ the fluid is contained in the interval $|z|<a$.
As time progresses the particles of a fluid move as indicated by narrow arrows pointing upward.
The particle located at $z$ at $t=0$ is shifted to the new position $z+\phi(z,t)$. 
A fluid volume shown as a narrow (red) horizontal rectangle occupies a segment $[z,z+dz]$ at time $t=0$.
At a later time $t>0$ this volume is displaced and occupies the segment $[z+\phi(z,t),z+dz+\phi(z+dz,t)]$.
The fluid volume expands (shrinks) if $\partial \phi(z,t)/ \partial z > (<) 0$.
The expansion (shrinkage) translates in the decrease (increase) in the density respectively.}
\label{Fig-phi}
\end{figure}

\textit{Hydrodynamic theory}. A hydrodynamic description is based on the existence of slow variables associated with locally conserved quantities such as number of particles, momentum and energy. The motion of the liquid is described by the Navier-Stokes equations which in Eulerian continuous field coordinates can be put in the form~\cite{LL} 
\begin{equation}\label{E-eq}
\partial_t(\rho v_j)=-\partial_i\Pi_{ij}-\rho\partial_jU,
\end{equation}
that guaranties the momentum conservation, which holds in ideal and nonideal liquids alike in the absence of confining potential $U$. The stress tensor of a two-dimensional fluid giving rise to the Navier-Stokes equation is~\cite{LL} 
\begin{equation}\label{Pi}
\Pi_{ij}=\delta_{ij}P+\rho v_iv_j-\zeta\delta_{ij}\partial_kv_k
-\eta(\partial_iv_j+\partial_jv_i-\delta_{ij}\partial_kv_k).
\end{equation}
Here $\eta,\zeta$ are the first (shear) and second (bulk) viscosities, and $P$ is pressure. In the following we focus on a 1D flow in $z$-direction of a two-dimensional liquid occupying the strip $|z|<a$, so that velocity vector field can be taken in the form $\bm{v}=v(z,t)\mathbf{e}_z$ and Eq.~(\ref{E-eq}) simplifies to 
\begin{equation}\label{NS-eq}
\rho(\partial_t v+v\partial_zv)=-\partial_z P-\rho \partial_z U+\partial_z(\eta\partial_zv).
\end{equation}  
When writing this equation we made use of the continuity equation $\partial_t\rho+\partial_z(\rho v)=0$ and employed standard assumption $\eta\gg\zeta$. For the purpose of our study it will be convenient to use a particle description of Navier-Stokes equation.
In this approach the coordinate $z$ labels an equilibrium position of a fluid particle, while its location at later time $t$, $z+ \phi(z,t)$ defines  the displacement field $\phi(z,t)$, see Fig.~\ref{Fig-phi}.
The density is $\rho(z+\phi(z,t),t) = \rho_0(z)/ ( 1 + \partial_z \phi(z,t))$, the velocity $v(z+\phi(z,t)) = \partial_t \phi(z,t)$, \cite{LL}.
The linearization of Eq.~\eqref{NS-eq} is equivalent to the linearization in $\phi$.
To the leading order, 
\begin{align}\label{leading}
\rho = \rho_0 + \delta \rho\, ,\,\, \delta \rho  = 
- (\rho_0 \phi)'  \, , \quad  v = \dot{\phi} \, ,
\end{align}
where $\rho_0$ is a stationary equilibrium density distribution, the notations $f' = \partial f/ \partial z$ and $\dot{f} = \partial f / \partial t$ are introduced and the pair of arguments $(z,t)$ common to all the functions is omitted.  
The parametrization \eqref{leading} of $\delta \rho$ and $v$ by a single displacement field automatically satisfies the linearized continuity equation, $\delta \dot{\rho} + (\rho_0 \dot{\phi})' = 0$.
The concept of the displacement field is further illustrated in Fig.~\ref{Fig-phi}. For the solutions of the form $\phi(z,t) = e^{i \omega t} \chi'(z)$ the equation \eqref{NS-eq} in the parameterization, \eqref{leading} reads, see App.~\ref{App}:
\begin{equation}\label{normal-modes-eq}
\omega^2\chi=-v^2_s\chi''+U'\chi' -i\omega\int^{z}_{z_0}\frac{dz}{\rho_0}(\nu \rho_0 \chi'')',
\end{equation}
where $\nu(z)=\eta(z)/\rho_0(z)$ is the kinematic viscosity, and
\begin{align}\label{v_s}
v_s=\sqrt{\partial P/\partial\rho_0}
\end{align}
has a meaning of a local speed of sound that depends on $z$ only through the equilibrium density $\rho_0$. We show that the results are independent on arbitrary $z_0$.

The spectrum of collective excitations and their decay rates as given by Eq.~\eqref{normal-modes-eq} depend on the details of the confining potential and dependence of the viscosity on density. 
Here for definiteness we consider the confining potential per unit mass with a weak quartic anharmonicity,
\begin{equation}\label{U}
U=\frac{\omega^2_\perp z^2}{2}+\frac{\epsilon z^4}{4m}+\delta U,
\end{equation}
where $m$ is the mass of an individual particles, and the constant $\delta U\propto\epsilon$ is added in such a way that the spatial extent of the liquid stays the same as for $\epsilon=0$. In other words $\delta U+\epsilon a^4/4m=0$. This choice is not obligatory, yet convenient in the further calculations. 
The anharmonicity $\propto\epsilon$ modifies both the first and the second term in the right-hand-side of Eq.~(\ref{normal-modes-eq}). Indeed, $v^2_s(z)=2\pi \rho(z)/m^3=\omega^2_\perp(a^2-z^2)/2$, valid for the parabolic confinement, acquires a correction $\delta v^2_s(z)=-\epsilon(z^4-a^4)/4m$.  
With these observations in mind we multiply Eq.~(\ref{normal-modes-eq}) by $2/\omega^2_\perp$, rescale coordinates $z\to z/a$ and introduce $\lambda^2_\omega=2\omega^2/\omega^2_\perp$ to find 
\begin{equation}\label{eigenmodes}
\lambda^2_\omega\chi+(1-z^2)\chi''-2z\chi'+V_\epsilon+V_\nu=0.
\end{equation}
The perturbation term due to anharmonicity can be cast in the form 
\begin{equation}\label{V1}
V_\epsilon=-\frac{\epsilon a^2}{2m\omega^2_\perp}[ (z^4-1)\chi ']'.
\end{equation}
The generalization of Eq.~\eqref{V1} to arbitrary confining potential is accomplished by replacing $[(z^4-1)\chi']'\to [(U(z)-U(a))\chi']'$. Consequently, our results for the depolarization shift are straightforward to modify for arbitrary confinement. 
We emphasize that $V_\epsilon$ is Hermitian. 
This property guaranties that the anharmonicity alone causes only the frequency shift but no dissipation. We argue below that only the combination of the anharmonicity and viscosity leads to the dissipation of the Kohn mode. The exact $z$-dependence of the viscosity $\nu(z)$ is specific to the model. This however only influences the numerical prefactors in the final results and we take the expression for the viscosity from the theory of Fermi liquids, $\eta\propto v_F\rho\ell$, where mean fee path is $\ell\propto v_FE_F/T^2$ with $v_F$ and $E_F$ being Fermi velocity and energy, respectively. 
For this case $\nu(z)=C\rho^2(z)/m^5T^2$ where $C$ is the numerical factor of the order of unity~\cite{Abrikosov}. In the above specified dimensionless notations this results in 
\begin{equation}\label{V2}
V_\nu=-i\lambda_\omega B\int^{z}_{z_0}\frac{dz}{1-z^2}[(1-z^2)^3  \chi'']',
\end{equation}
where we have introduced $B=\sqrt{2}\nu_0/a^2\omega_\perp$ with $\nu_0=\nu(z=0)$. We proceed with the analysis of Eq.~(\ref{eigenmodes}). 

\textit{Results}. As the first step let us discuss the eigenmodes of an ideal fluid confined to a harmonic trap. For that purpose we neglect anharmonicity and interaction effects implicit in the viscosity term of Eq.~(\ref{eigenmodes}), i.e. we set $V_{\epsilon} =V_{\eta}=0$. 
What remains is familiar Legendre equation and we therefore immediately read off its solutions $\lambda^2_{\omega_n}=2\omega^2_n/\omega_\perp=n(n+1)$ with $n=0,1,2,\ldots$ so that the eigenfrequencies are 
\begin{equation}\label{omega-n}
 \omega_n=\omega_\perp\sqrt{\frac{n(n+1)}{2}}. 
\end{equation}
The $n=0$ gives an equilibrium since $\chi_{n=0}=const$ this velocity is identically zero. The $n=1$ is a Kohn mode $\omega_1=\omega_\perp$. We thus found the whole hierarchy of eigenoscillations, they are Legendre polynomials 
\begin{equation}\label{chi}
\chi_n(z)=\sqrt{\frac{2n+1}{2}}P_n(z)
\end{equation} 
and the related velocity fields are $v_n(z)=\partial_zP_n$. 
Remarkably, the same spectrum of collective modes as \eqref{omega-n} was obtained recently for the longitudinal oscillations of the 1D Coulomb chains \cite{Morigi}.
In this systems the collective behavior sets in due to the long range Coulomb forces rather than collisions.

Next we discuss the significance of perturbation terms on the spectrum of collective modes. Let us first consider anharmonicity alone, i.e. we set $V_{\eta}=0$ and $V_{\epsilon}\neq 0$ in Eq.~\eqref{eigenmodes}. As $V_{\epsilon}$ is Hermitian the spectrum remains real.
As a result all the eigenmodes remain undamped as expected in the absence of collisions.
To find the depolarization shift of the Kohn mode, $\delta^{(1)} \omega_1$ to the leading order in $\epsilon$ we write $\lambda^2_{\omega_1}=2+\delta^{(1)} \lambda^2_{\omega_1}$, where the correction term is found from the first order perturbation theory with unperturbed solutions given by Eq.~\eqref{chi},
\begin{equation}
\delta^{(1)} \lambda^2_{\omega_1}=\frac{3\epsilon a^2}{4m\omega^2_\perp}
\langle P_1(z)|\partial_z[ (z^4-1)\partial_z]|P_1(z)\rangle.
\end{equation}
The matrix element gives a factor of $8/5$ which eventually translates into the correction to the Kohn mode frequency 
\begin{equation}\label{dep_shift}
\delta^{(1)}\omega_1=\frac{3\epsilon a^2}{10m\omega_\perp}.
\end{equation}
We have checked that Eq.~\eqref{dep_shift} agrees with the result obtained by the method of moments suggested in Ref.~\cite{Pantel}. Note however that the depolarization shift in the collisioneless regime differ from Eq.~\eqref{dep_shift} by a nonuniversal numerical prefactor. 
For instance, for the contact interaction we obtain by direct perturbation theory in Fermions $\delta^{(1)}\omega_1=3\epsilon a^2/5m\omega_\perp$, \cite{Iqbal2}. Corrections to other eigenfrequencies can be computed in the same fashion with the result \eqref{dep_shift} up to a numerical coefficient. 

Before considering the generic case, it is instructive to verify the Kohn's theorem within our hydrodynamic approach. It amounts to the statement that $\chi_1(z) \propto z$ remains the solution of Eq.~(\ref{eigenmodes}) with the frequency $\omega_1 = \omega_{\perp}$ even for $V_{\eta} \neq 0$ provided only $V_{\epsilon}=0$. This in turn will be proven once we show that for any non-negative $n$, $\langle P_1  | V_{\eta} | P_n \rangle = \langle P_n  | V_{\eta} | P_1 \rangle =0$. Clearly $V_{\eta} | P_1 \rangle =0$ as $[P_1(z)]''=0$. On the other hand from Eq.~(\ref{normal-modes-eq}) it follows that 
\begin{align}
\!\!\!\langle P_1|V_\nu| P_n\rangle \!\!\propto\!\!\int^{1}_{-1}\!\!\!\!d\bar{z} P_1(\bar{z})\!\!\! \int^{\bar{z}}_{z_0}\!\!\frac{dz}{\rho_0(z)}\{\nu(z) \rho_0(z) [P_n(z)]'\}'.
\end{align}
Realizing that $P_1(\bar{z})=-\partial_z \rho_0(\bar{z})/2$ and integrating by parts one concludes that $\langle P_1|V_\nu| P_n\rangle=0$ for all $n$ thus proving the Kohn theorem in the present context.

It follows that the Kohn mode may acquire a finite life-time only when the anharmonicity is included.
Yet the perturbation $V_{\epsilon}$ is Hermitian and by itself is insufficient.
We therefore consider both perturbations, and look for the corrections that are first order in each of the two. In the second order perturbation theory such a correction is of the form     
\begin{eqnarray}
&&\hskip-.25cm\delta^{(2)}\lambda^2_{\omega_1}=-i\lambda_\omega B\sum^{\infty}_{n=1}\frac{\sqrt{3/2}\sqrt{(2n+1)/2}}{2-n(n+1)}\nonumber\\
&&\hskip-.25\times[\langle P_1|V_\nu| P_n\rangle\langle P_n|V_\epsilon| P_1\rangle+
\langle P_1|V_\epsilon| P_n\rangle\langle P_n|V_\nu| P_1\rangle].
\end{eqnarray}
As we saw above both terms in this equation are zero as the Kohn mode does not couple to any other mode by a viscosity term $V_{\eta}$, and $\delta^{(2)}\lambda^2_{\omega_1}=0$.

Inevitably we have to consider the third order corrections to $\lambda_{\omega_1}$. 
The third order correction to the energy $E_l$ of a state $|l\rangle$ is 
\begin{eqnarray}
\delta^{(3)} E_l=\sum_{k,m\neq l}\frac{\langle l|V|m\rangle\langle m|V|k\rangle\langle k|V|l\rangle}{(E_m-E_l)(E_k-E_l)}\nonumber\\
-\langle l|V|l \rangle\sum_{m\neq l}\frac{\langle l|V|m\rangle\langle m|V|l\rangle}{(E_m-E_l)^2}. 
\end{eqnarray}
To apply this expression to our problem we identify $|l\rangle=P_1$ as a Kohn mode and $V=V_\epsilon+V_\nu$. We observe that for the perturbation terms given by Eqs.~(\ref{V1}) and (\ref{V2}) the following properties of matrix elements hold $\langle P_1|V_\nu|P_n\rangle=\langle P_n|V_\nu|P_1\rangle=\langle P_1|V_\epsilon|P_1\rangle=0$ for all $n$. Furthermore, for the quartic anharmonicity under consideration the only nonzero off-diagonal matrix elements are $\langle P_1|V_\epsilon|P_3\rangle=\langle P_3|V_\epsilon|P_1\rangle$ with the rest of matrix elements $\langle P_n|V_\epsilon|P_1\rangle=0$ for $n\neq1,3$. We thus have, accounting for all the normalization factors of eigenoscillation modes (\ref{chi}) 
\begin{equation}
\delta^{(3)}\lambda^2_{\omega_1}=\frac{147}{8(\lambda^2_{\omega_1}-\lambda^2_{\omega_3})^2}\langle P_1|V_\epsilon|P_3\rangle
\langle P_3|V_\nu|P_3\rangle
\langle P_3|V_\epsilon|P_1\rangle.
\end{equation}
Using the explicit expressions (\ref{V1}) and (\ref{V2}) we find for the matrix elements 
\begin{eqnarray}
&&\hskip-.25cm\int^{1}_{-1}d\bar{z}P_3(\bar{z})\!\!\int^{\bar{z}}_{z_0}\frac{dz}{1-z^2}\{(1-z^2)^3[P_3(z)]''\}'=-\frac{40}{21},\\
&&\hskip-.25cm\int^{1}_{-1}dz P_1(z)\{(z^4-1)[P_3(z)]'\}'=\frac{8}{5},
\end{eqnarray}
and eventually
\begin{equation}
\delta^{(3)}\lambda^2_{\omega_1}=i\lambda_\omega\frac{28\sqrt{2}}{125}\frac{\nu_0}{a^2\omega_\perp}
\left(\frac{\epsilon a^2}{m\omega^2_\perp}\right)^2.
\end{equation}
This result enables us to find imaginary part of the Kohn mode $\omega_1=\omega_\perp+\delta\omega_1+i\tau^{-1}_1$, which corresponds to its attenuation with the rate 
\begin{equation}\label{tau-K}
\tau^{-1}_1\simeq\frac{\nu_0\epsilon^2a^2}{m^2\omega^4_\perp}\simeq\left(\frac{\delta\omega_1}{\omega_\perp}\right)^2\frac{\nu_0}{a^2},
\end{equation}
where we omitted numerical factors of order unity for brevity. This expression constitutes the main result of our work and has straightforward interpretation. The higher excitations modes not protected by the Kohn theorem decay with the rate $\sim\nu_0/a^2$. As $\nu_0$ has dimensions of the diffusion coefficient, this is the typical rate of the momentum relaxation. The ratio $(\delta\omega_1/\omega_\perp)^2$ is the probability of finding the system in higher modes.    

\textit{Discussions}. Hydrodynamic description requires short equilibration length $\ell$. Thus validity of our theory is limited by the condition $\ell\ll a$, which imposes certain restriction on temperature. Specifically, for the Fermi liquids $\ell=v_F\tau_{ee}$ is determined by collisions with the typical rate $\tau^{-1}_{ee}\sim T^2/E_F$. Since $\omega_\perp\sim v_F/a$, hydrodynamic regime is realized at temperatures $T>T_h$ above the crossover scale $T_h\sim\sqrt{\omega_\perp E_F}\sim E_F/\sqrt{N}$, where $N$ is the number of occupied sub-bands of the transversal quantization. It also follows that with necessity hydrodynamics requires $T\gg\omega_\perp$. While this inequality is reasonably satisfied for the cold gases that are confined by a very shallow potential, it obviously breaks in the ultra-cold limit where collisionless regime prevails. In the latter case attenuation coefficient of the Kohn mode is expected to follow quadratic temperature dependence $\tau^{-1}_{1}\propto \alpha T^2/E_F$ based on the Pauli principle and phase space restrictions argument, whereas in the hydrodynamic regime $\tau^{-1}_{1}\propto 1/T^2$ in accordance with Eq.~(\ref{tau-K}). The nonmonotonic temperature dependence of the decay rate has been observed experimentally~\cite{Riedl}.  

Our hydrodynamic approach has interesting parallels with the Luttinger liquid description of collective modes in confined inhomogeneous 1D gases ~\cite{Citro}. The eigenvalue equation for the normal eigenmodes in that case, analogous to our Eq.~(\ref{normal-modes-eq}), is given by  
\begin{equation}\label{eigenmodes-LL}
-\omega^2_n\chi_n(z)=v(z)K(z)\partial_z\left(\frac{v(z)}{K(z)}\partial_z\chi_n(z)\right)\, ,
\end{equation}
where Luttinger liquid interaction parameter satisfies the relation $v(z)K(z)=\pi \rho(z)/m^2$. This equation is supplemented by the boundary condition $\chi_n(\pm a)=0$ and normalization condition $\int^{a}_{-a} dz \chi_j(z)\chi_j(z)/v(z)K(z)=\delta_{ij}$. For the particular choice of $v(z)=v_0\sqrt{1-z^2/a^2}$ and $K(z)=K_0(1-x^2/a^2)^\gamma$ the solutions $\chi_n(z)$ are obtained in terms of Gegenbauer polynomials with the spectrum of excitations $\omega^2_n=(v_0/a)^2(n+1)(n+2\gamma+1)$~\cite{Citro,Petrov1,Stringari}. In the model of $\gamma=2$, the problem simplifies to the case of Legendre polynomials~\cite{Petrov2} with the spectrum of excitations analogous to our result (\ref{omega-n}). Another interesting limit is $\gamma=0$, which corresponds to the case of the Tonks-Girardeau gas, where the Gegenbauer polynomials reduce to Chebyshev polynomials. Inclusion of dissipative terms into Eq.~(\ref{eigenmodes-LL}) requires consideration of corrections to Luttinger liquid model which account for the inelastic scattering of bosons and ultimately describe equilibration processes. As recently shown such generalization is possible both in the limit of weak \cite{AL} and strong \cite{Matveev} interactions and application of this formalism to the problem of decay of collective modes is an interesting problem for future research. Along this rout one may hope to find a unified description, which interpolates between the quantum~\cite{Matveev} and classical~\cite{Andreev} hydrodynamic regimes of Luttinger liquids, and which is broadly applicable for arbitrarily strong interactions.   

\textit{Acknowledgment}. We thank M. Raikh for discussions that attracted our attention to this interesting problem. We indebted to E. Bettelheim for pointing out to us importance of certain correction terms to Navier-Stokes equations in the inhomogeneous geometry. We also thank M. Dykman for the discussion of results. This work at MSU (A.L.) was supported by NSF Grant No. DMR-1401908.  
A.I. and M.K. are grateful to the University of Iowa for the support.

\begin{appendix}
\section{Derivation of equation~\eqref{normal-modes-eq}}
\label{App}
Here we derive the Eq.~\eqref{normal-modes-eq} of the main text.
In the parametrization \eqref{leading}, the left hand side of Eq.~\eqref{NS-eq} takes the form, 
\begin{align}\label{LHS}
\rho(\partial_t v+v\partial_zv) \approx \rho_0 \ddot{\phi}\, .
\end{align}
To linearize the right hand side of Eq.~\eqref{NS-eq} we note that the pressure is fixed by the density via the equation of state such that 
\begin{align}
P(z,t) = P[\rho(z,t)] \approx P[\rho_0(z)] - v_s^2[\rho_0(z)] (\rho_0 \phi)'\, ,
\end{align}
where the velocity $v_s$ is defined in Eq.~\eqref{v_s}.
At equilibrium, Eq.~\eqref{NS-eq} yields 
\begin{align}\label{cond}
v_s^2 \rho_0' = -\rho_0 U'
\end{align}
We have therefore,
\begin{align}
-P'-\rho U' \approx 
[ v_s^2 (\rho_0 \phi)']' +
(\rho_0 \phi)' U'
 \end{align}
Writing $(\rho_0 \phi)' U' =[ \rho_0 \phi U']' - \rho_0 \phi U''$ and using \eqref{cond} we obtain,
\begin{align}
-P'-\rho U' \approx 
[ v_s^2 \rho_0 \phi ']' 
- (\rho_0 \phi) U''
 \end{align}
Writing 
\begin{align*}
[ v_s^2 \rho_0 \phi ']'  =\rho_0 [ v_s^2 \phi ']' +  \rho_0' [ v_s^2 \phi ']
\end{align*}
and using \eqref{cond} again we obtain
\begin{align}\label{first}
-P'-\rho U' \approx 
\rho_0 [ v_s^2 \phi ']' -  
\rho_0 \phi ' U'
- \rho_0 \phi U''\, .
 \end{align}
The third, viscosity term on the right hand side of Eq.~\eqref{NS-eq} reads
\begin{align}\label{third}
\partial_z(\eta\partial_z v) = [\eta \dot{\phi}']'
\end{align}
Substituting Eqs.~\eqref{LHS}, \eqref{first} and \eqref{third} in Eq.~\eqref{NS-eq} of the main text we obtain
\begin{align}
\ddot{\phi} = 
[ v_s^2 \phi ']' -  
 \phi ' U'
- \phi U'' + \rho_0^{-1} [\eta \dot{\phi}']'
\end{align}
For the solutions of the form $\phi(z,t) = e^{i \omega t} \chi'(z)$ we obtain the equation,
\begin{align}\label{interim}
 - \omega^2 \chi' = 
[ v_s^2 \chi'']' -  
\chi'' U'
- \chi' U'' +  (- i \omega)\rho_0^{-1} [\eta \chi'']'
\end{align}
Integration of Eq.~\eqref{interim} over $z$ yields Eq.~\eqref{normal-modes-eq}.

\end{appendix}


\begin{thebibliography}{99}

\bibitem{Review-1}
V.~V.~Deshpande, M.~Bockrath, L.~I.~Glazman, A.~Yacoby, Nature \textbf{464}, 209 (2010). 

\bibitem{Review-2}
M.~A.~Cazalilla, R.~Citro, T.~Giamarchi, E.~Orignac, and M.~Rigol, 
Rev. Mod. Phys. \textbf{83}, 1405 (2011).

\bibitem{Review-3}
A.~Imambekov, T.~L.~Schmidt, and L.~I.~Glazman,
Rev. Mod. Phys. \textbf{84}, 1253 (2012). 

\bibitem{Haldane} 
F.~D.~M.~Haldane, J. Phys. C: Solid State Phys., \textbf{14}, 2585 (1981).

\bibitem{Stone} 
M.~Stone, \textit{Bosonization}, (World Scientific Publishing Co., 1994).

\bibitem{Gogolin}
A.~O.~Gogolin, A.~A.~Nersesyan, and A.~M.~Tsvelik, \textit{Bosonization and strongly correlated
systems}, (Cambridge University Press, 1998).

\bibitem{Giamarchi}
T. Giamarchi, \textit{Quantum Physics in One Dimension}, (Claredon Press, Oxford, 2003).

\bibitem{Mattis}
D.~C.~Mattis, \textit{The Many-Body Problem: An Encyclopedia of Exactly Solved Models in One Dimension}, (World Scientific Publishing, 1992).

\bibitem{Sutherland} 
B.~Sutherland, \textit{Beautiful models: 70 years of exactly solved quantum many-body problems}, (World Sci. Pub., 2004).

\bibitem{Kohn}
W. Kohn, Phys. Rev. \textbf{123}, 1242 (1961). 

\bibitem{Dobson}
J.~F.~Dobson, Phys. Rev. Lett. \textbf{73}, 2244 (1994).

\bibitem{Kinast}
J.~Kinast, S.~L.~Hemmer, M.~E.~Gehm, A.~Turlapov, and J.~E.~Thomas, 
Phys. Rev. Lett. \textbf{92}, 150402 (2004).

\bibitem{Bartenstein}
M.~Bartenstein, A.~Altmeyer, S.~Riedl, S.~Jochim, C.~Chin, J.~Hecker Denschlag, and R.~Grimm, 
Phys. Rev. Lett. \textbf{92}, 203201 (2004).

\bibitem{Altmeyer}
A.~Altmeyer, S.~Riedl, C.~Kohstall, M.~J.~Wright, R.~Geursen, M.~Bartenstein, C.~Chin, J.~Hecker Denschlag, and R.~Grimm, Phys. Rev. Lett. \textbf{98}, 040401 (2007).

\bibitem{Wright}
M.~J.~Wright, S.~Riedl, A.~Altmeyer, C.~Kohstall, E.~R.~Sanchez Guajardo, J.~Hecker Denschlag, and R.~Grimm, Phys. Rev. Lett. \textbf{99}, 150403 (2007).

\bibitem{Khodas}
M.~Khodas, M.~Pustilnik, A.~Kamenev, and L.~I.~Glazman, 
Phys. Rev. B \textbf{76}, 155402 (2007).

\bibitem{Barak}
G.~Barak, H.~Steinberg, L.~N.~Pfeiffer, K.~W.~West, L.~Glazman,
F.~von Oppen, and A.~Yacoby, Nat. Phys. \textbf{6}, 489 (2010).

\bibitem{Karzig}
T.~Karzig, L.~I.~Glazman, and F.~von Oppen, Phys. Rev. Lett. \textbf{105}, 226407 (2010).

\bibitem{Micklitz}
T.~Micklitz and A.~Levchenko, Phys. Rev. Lett. \textbf{106}, 196402 (2011).

\bibitem{Levchenko}
A.~Levchenko, Phys. Rev. Lett. \textbf{113}, 196401 (2014).

\bibitem{Brey} 
L.~Brey, N.~F.~Johnson, and B.~I.~Halperin, Phys. Rev. B \textbf{40}, 10647 (1989).

\bibitem{Iqbal} 
A.~Iqbal and M.~Khodas, Phys. Rev. B \textbf{90}, 155439, (2014).

\bibitem{Drexler} 
H.~Drexler, W.~Hansen, J.~P.~Kotthaus, M.~Holland, and S.~P.~Beaumont, 
Phys. Rev. B \textbf{46}, 12849(R) (1992).

\bibitem{Wendler} 
L.~Wendler and R.~Haupt, Phys. Rev. B \textbf{52}, 9031 (1995).

\bibitem{Pantel} 
P.~A.~Pantel, D.~Davesne, S.~Chiacchiera, and M.~Urban, Phys. Rev. A \textbf{86}, 023635 (2012).

\bibitem{Schneider} 
S.~Schneider and G.~J.~Milburn, Phys. Rev. A \textbf{65}, 042107 (2002).

\bibitem{Riedl} 
S.~Riedl, E.~R.~Sanchez Guajardo, C.~Kohstall, A.~Altmeyer, M.~J.~Wright, J.~Hecker Denschlag, R.~Grimm, G.~M.~Bruun, and H.~Smith, Phys. Rev. A \textbf{78}, 053609 (2008).

\bibitem{LL}
L.~D.~Landau and E.~M.~Lifshitz, \textit{Fluid Mechanics} (Pergamon Press, Oxford, 1987).


\bibitem{Morigi}
G.~Morigi and S.~Fishman, Phys. Rev. Lett. \textbf{93}, 170602 (2004);
G.~Morigi and S.~Fishman, Phys. Rev. E \textbf{70}, 066141 (2004). 

\bibitem{Abrikosov}
A.~A.~Abrikosov and I.~M.~Khalatnikov, Rep. Prog. Phys. \textbf{22}, 329 (1959).

\bibitem{Iqbal2}
A. Iqbal, A. Levchenko, and M. Khodas, unpublished.

\bibitem{Citro}
R.~Citro, S.~De Palo, E.~Orignac, P.~Pedri, and M.-L.~Chiofalo, New J. Phys. \textbf{10}, 045011 (2008).

\bibitem{Stringari}
C.~Menotti and S.~Stringari, Phys. Rev. A \textbf{66}, 043610 (2002).

\bibitem{Petrov1}
D.~Petrov, D.~Gangardt, and G.~Shlyapnikov, J. Physique IV \textbf{116}, 5 (2004). 

\bibitem{Petrov2}
D.~S.~Petrov, G.~V.~Shlyapnikov, and J.~T.~M.~Walraven, 
Phys. Rev. Lett. \textbf{85}, 3745 (2000).

\bibitem{AL}
A.~Levchenko, T.~Micklitz, J.~Rech, and K.~A.~Matveev, Phys. Rev. B \textbf{82}, 115413 (2010).

\bibitem{Matveev}
W.~DeGottardi and K.~A.~Matveev, arXiv:1412.0693.

\bibitem{Andreev}
A.~V.~Andreev, S.~A.~Kivelson, and B.~Spivak, Phys. Rev. Lett. \textbf{106}, 256804 (2011).

\end{thebibliography}
\end{document}